\documentclass[pdflatex,sn-mathphys-num]{sn-jnl}

\usepackage{graphicx}%
\usepackage{multirow}%
\usepackage{amsmath,amssymb,amsfonts}%
\usepackage{amsthm}%
\usepackage{mathrsfs}%
\usepackage[title]{appendix}%
\usepackage{xcolor}%
\usepackage{textcomp}%
\usepackage{manyfoot}%
\usepackage{booktabs}%
\usepackage{algorithm}%
\usepackage{algorithmicx}%
\usepackage{algpseudocode}%
\usepackage{listings}%
\usepackage{bm}
\usepackage{color, soul}

\usepackage{amsmath,amssymb,amsfonts}
\usepackage{bm}
\usepackage{color, soul}
\usepackage{graphicx}
\usepackage{float}
\usepackage{wrapfig}
\usepackage{verbatim}
\usepackage[english]{babel}
\usepackage{amstext, mathrsfs, textcomp}
\usepackage{multirow}
\usepackage{siunitx}
\usepackage{enumerate}
\usepackage{enumitem}
\usepackage{epstopdf}
\usepackage{tabularx}
\makeatletter

\usepackage{physics}
\usepackage{upgreek}

\sethlcolor{yellow}

\graphicspath{{Figures/}}

\begin{document}

\title[Metasurface-Enhanced Mid-Infrared Imaging Spectroscopy with Broadband Quantum Cascade Lasers]{Metasurface-Enhanced Mid-Infrared Imaging Spectroscopy with Broadband Quantum Cascade Lasers}


\author[1]{\fnm{Ivan} \sur{Sinev}}

\author[2]{\fnm{Alessio} \sur{Cargioli}}

\author[2]{\fnm{Diego} \sur{Piciocchi}}

\author[1]{\fnm{Felix Ulrich} \sur{Brikh}}

\author[2]{\fnm{Jérôme} \sur{Faist}}

\author*[1]{\fnm{Hatice} \sur{Altug}}\email{hatice.altug@epfl.ch}

\affil*[1]{\orgdiv{Institute of Bioengineering}, \orgname{\'Ecole Polytechnique F\'ed\'erale de Lausanne (EPFL)}, \orgaddress{\city{Lausanne}, \postcode{1015}, \country{Switzerland}}}

\affil[2]{\orgdiv{Institute for Quantum Electronics}, \orgname{ETH Zurich}, \orgaddress{\city{Zurich}, \postcode{CH-8093}, \country{Switzerland}}}


\abstract{Mid-infrared (mid-IR) spectroscopy offers unparalleled opportunities in sensing through chemically specific detection of molecular absorption fingerprints. Yet, its practical applications are limited by the weak light-matter interaction in the mid-IR range and low brightness of mid-IR light sources. Surface-enhanced infrared absorption (SEIRA) spectroscopy addresses the sensitivity limitations by leveraging resonant photonic structures, in particular, plasmonic and frequency-selective dielectric metasurfaces. However, current implementations of SEIRA approach mainly rely on complex instruments and scanning components such as Fourier-transform infrared spectroscopy and tunable external cavity quantum cascade lasers (EC QCLs). Here, we present a compact and high-throughput imaging-based SEIRA platform that combines broadband gradient metasurfaces with a radiofrequency-modulated QCL that generates remarkably broad instantaneous emission spectrum (250~cm$^{-1}$) covering absorption bands of multiple distinct molecular vibrational modes. By matching the resonance spectrum of the compact (1~mm$^2$) broadband gradient metasurface with the laser emission projected on its surface through a dispersive element, we ensure that every QCL spectral component is uniquely addressed for an efficient targeted enhancement of the electromagnetic field. This enables us to use a low-cost and room-temperature mid-IR camera, acquiring in a single frame the enhanced absorption signatures of analytes deposited on the metasurface as a barcode image, thus reducing the acquisition time by up to 3 orders of magnitude compared to the FTIR and EC QCL based measurements. Eliminating the need for tunable light sources, bulky spectrometers, and expensive low-temperature detectors, our approach enables high-throughput, miniaturized, and highly specific molecular diagnostics for diverse chemical and biological applications.}

\maketitle

\section*{Introduction}

Mid-infrared (mid-IR) spectroscopy finds use in numerous scientific and industrial applications, including food and environmental monitoring\cite{wilson1999mid,willer2006near}, medical diagnostics\cite{wang2008application,de2018applications}, chemical and biological sensing\cite{haas2016advances,caixeta2023monitoring}. Its ability to probe molecular vibrations provides highly specific spectral signatures, the so-called molecular fingerprints, making it an indispensable technique for material identification and composition analysis. However, the giant mismatch between the mid-IR wavelengths and the characteristic sizes of molecules leads to extremely low absorption cross section of the molecular species, which requires large quantities of analytes and long acquisition times to accumulate sufficient signal. Surface-enhanced infrared absorption spectroscopy (SEIRA) mitigates these issues through engineering of the enhanced local electromagnetic fields enabled by resonant photonic structures\cite{Neubrech2017ChemRev,Herpin2023AdvMAt,Paggi2023NatCom}, targeting the relevant mid-IR absorption bands. The amplified light-matter interaction arising from efficient coupling of optical modes with molecular vibrations provides orders of magnitude improvements to the detected absorption signal. 

Plasmonics are the conventional choice for achieving the strong local field enhancement required for SEIRA. Various designs based on plasmonic mid-IR nanoantennas have promoted zepto-mole limits of detection for protein species~\cite{Adato2009Nov}, precise discrimination of molecular conformations~\cite{Kavungal2023ScienceAdvances}, and simultaneous sensing in multiple spectral ranges enabled by self-similar geometries~\cite{Gottheim2015ACSNano,Rodrigo2018NatComm}. More recently, metasurfaces based on high refractive index dielectric materials emerged as a viable alternative in SEIRA. Their distinct feature is low ohmic losses, which results in potentially much higher optical mode Q-factors as compared to plasmonics. The associated strong but narrow-band field enhancement enables extremely specific targeting of molecular vibrational modes, which facilitates reaching strong coupling regimes\cite{Guddala2021Oct,Biswas2024Jul,Richter2024Jun,Adi2024Nov}. Yet, for many applications, including spectroscopic sensing, broadband operation is desirable. Such functionality was realized with high-Q pixelated metasurfaces\cite{Tittl2018Science} that offered discrete sampling of a broad (1350-1750~cm$^{-1}$) mid-IR spectral range, further converted into a barcode-like spatial absorption map for imaging-based sensing. Most recently, resonance gradient metasurface were proposed to provide continuous spectral coverage over a broad bandwidth as well as a smaller device footprint \cite{Jangid2023AdvMat,Richter2024Jun,aigner2024continuous}.

\begin{figure}[ht!]
    \centering
    \includegraphics[width=\textwidth]{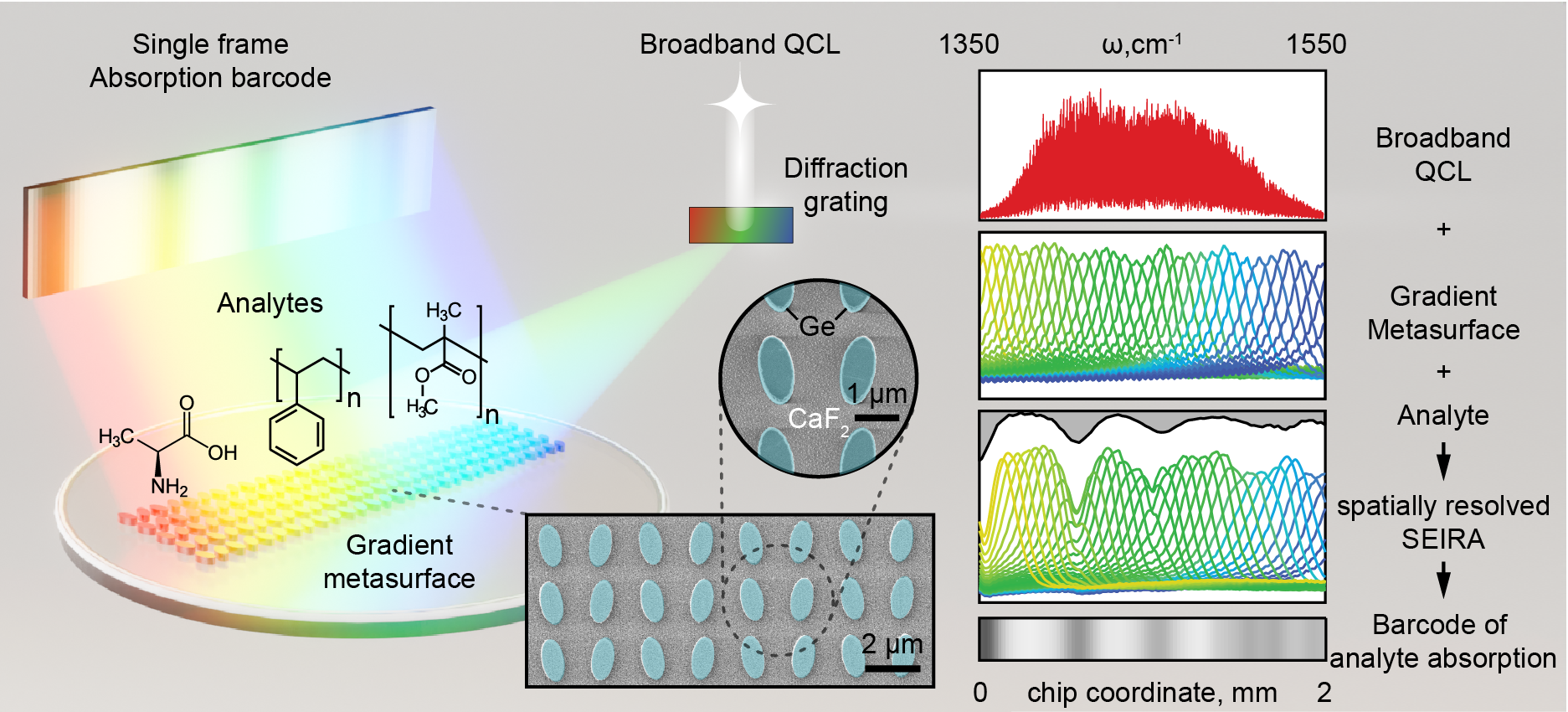}
    \caption{\textbf{Illustration of rapid imaging-based SEIRA spectroscopy enabled by the synergy of resonance gradient metasurfaces and RF-modulated QCL generating broad instantaneous emission spectrum.} Laser emission is projected on the metasurface through a dispersive element, providing spatially resolved SEIRA signal. The gradient of the metasurface resonance is tailored to match the spectral component of the broadband laser addressing each location on the chip. The reflectivity of the resonance gradient metasurface is then captured in a single frame of a room-temperature mid-IR imaging camera, allowing to detect the absorption signature of analytes deposited on the metasurface in a barcode-like format. The insets show the false color SEM images of the gradient metasurface.}
    \label{fig:Intro}
\end{figure}

Despite these advances, standard mid-IR spectroscopy and SEIRA still rely on complex measurement instrumentation, such as Fourier transform infrared (FTIR) spectrometers, which poses a significant challenge for broader implementation of both techniques. One of the major limiting factors is the inherently low brightness of mid-IR light sources. FTIRs address this issue by detecting the integral signal of thermal sources (globars) that provide  broadband emission covering range from 600 to 2500~cm$^{-1}$. Nonetheless, even under these conditions, highly sensitive detectors -- such as HgCdTe (mercury cadmium telluride, MCT) operating at cryogenic temperatures -- are necessary. 

The development of quantum cascade lasers (QCLs) presents a promising alternative to traditional mid-IR sources. QCLs is an inherently narrowband (few MHz) source offering high-intensity, coherent radiation, which significantly improves the spectral resolution and simplifies detection. Given the importance of broadband mid-IR spectroscopy, QCL architectures that are capable of addressing broader spectral ranges gained particular attention. State of the art external cavity (EC) QCLs\cite{hugi2009external,hugi2010external} were able to reach tuning ranges of up to several hundreds of cm$^{-1}$ with mechanically scanned components\cite{gmachl_ultra-broadband_2002}. Combination of tunable EC QCLs with pixelated SEIRA metasurfaces that provide spatially encoded spectral information facilitated a shift toward spectrometerless, imaging-based mid-IR characterization techniques\cite{hasenkampf2015surface,Tittl2018Science,aigner2024continuous}.  However, the use of EC entails extended acquisition times dictated by the moving parts of the laser tuning system, which require a higher degree of mechanical stability of the system and in general a larger setup footprint \cite{mroziewicz_external_2008}. Frequency comb (FC) QCLs\cite{faist2016quantum} differ from the EC QCL by the fact that they provide instantaneous broad spectrum. As a result, they have recently emerged as a promising on-chip and compact source  to perform rapid mid-IR analysis over a wide bandwidth. In particular, dual-comb spectroscopy techniques offer short ($\mu$s) acquisition times and no moving parts\cite{villares2014dual}, significantly improving the measurement reliability. While the FC bandwidth is typically limited to $\sim 100~$cm$^{-1}$\cite{singleton_evidence_2018, heckelmann_quantum_2023}, it was recently proven that a strong radio-frequency modulation of a narrowband QCL can result in a remarkably broader spectral coverage. This modulation drives the device to operate in a quasi-coherent regime, where the typical linewidth of the laser is of the order of the modulation frequency \cite{Cargioli2024Mar} (hundreds of MHz) and the full spectrum is emitted in a single laser pulse. Importantly, with this approach it is possible to use the same active material employed in tunable EC-lasers but without requiring any moving parts and mechanical stabilization, which results in a platform that is highly appealing for broadband spectroscopy, including SEIRA.

In this paper, we propose a rapid imaging-based SEIRA spectroscopy technique enabled by the synergistic combination of broadband resonance gradient metasurfaces and an RF-modulated QCL generating remarkably broad instantaneous emission spectrum over 250~cm$^{-1}$ (Fig.~\ref{fig:Intro}). In our measurement configuration, we match the local resonance of a compact (footprint $\approx 1~\text{mm}^2$) gradient metasurface with the spectral dispersion of the laser radiation projected on it for an efficient targeted enhancement of the electromagnetic field. The broad emission range of our laser simultaneously covers absorption bands of multiple distinct molecular vibrational modes without any time-consuming mechanical tuning that is required in conventional EC QCL systems. This led us to use a low-cost and room-temperature mid-IR imaging camera to capture in a single frame the enhanced absorption signatures of analytes deposited on the gradient chip as a barcode image, improving the detection time by up to 3 orders of magnitude as compared to FTIR and EC QCL based systems. We envision the use our approach for rapid and specific diagnostics of diverse chemical compounds, leading to compact and low-cost mid-IR sensors for operation at remote settings.

\begin{figure}[ht!]
    \centering
    \includegraphics[width=0.8\textwidth]{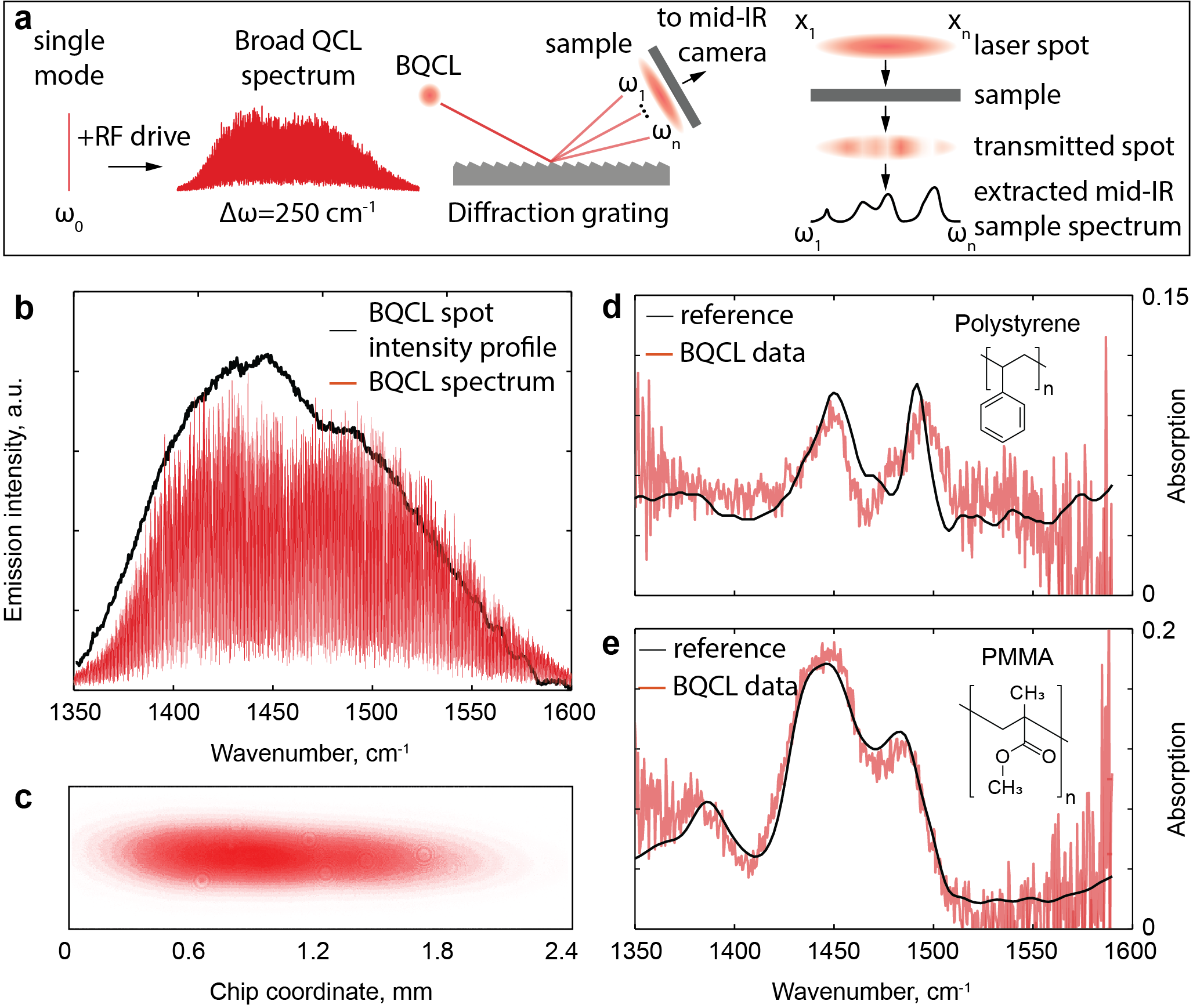}
    \caption{\textbf{Non-enhanced imaging spectroscopy with broadband quantum cascade laser} (a) Schematic of the broadband QCL imaging spectroscopy components. Left: quantum cascade laser is driven into broadband emission regime using radiofrequency drive. Center: broadband laser emission is projected on a sample surface through a dispersive element and then imaged with a mid-IR camera. Right: as each frequency component of BQCL addresses a specific coordinate on the sample surface, the spatial profile of the laser spot represents the sample transmission spectrum. (b) Spatial intensity profile of a projected laser spot (black curve) compared to the spectrum of BQCL emission measured with FTIR (red curve). (c) Image of the BQCL spot on the surface of CaF$_2$ chip recorded with a mid-IR camera. (d,e) Comparison of the absorption spectra of (d) polystyrene and (e) polymetylmetacrylate films on CaF$_2$ measured with FTIR (black curves) and extracted from the spatial intensity profiles of the transmitted BQCL beam (red curves).}
    \label{fig:spectroscopy}
\end{figure}

\section*{Results}
The broadband QCL (BQCL) device we employ is fabricated using the buried heterostructure technique\cite{Beck2001Dec} and emits at around 1430~cm$^{-1}$ when driven with DC current. We superimposed an amplified RF signal (29~dBm, 500~MHz) on DC current with a bias tee to drive the laser into a broadband emission regime\cite{Cargioli2024Mar} as illustrated in Fig.~\ref{fig:spectroscopy}a. The laser emission spectrum in this regime measured with FTIR is shown in Fig.~\ref{fig:spectroscopy}b and it features frequency-equidistant narrow emission lines with free spectral range of 0.5~cm$^{-1}$
. To establish the operation of the device without the use of FTIR, we focused the laser beam on the surface of a bare 0.5~mm thick CaF$_2$ chip through a reflective diffraction grating (see Methods for more details). In this configuration, each frequency component of the laser addresses a distinct location on the sample surface. Therefore, the intensity profile of the BQCL spot represents the envelope of the laser spectrum, slightly modified due to the finite focussing power of the imaging lenses. Figure~\ref{fig:spectroscopy}c shows an example of BQCL spot transmission captured with the room-temperature mid-IR camera (DataRay IR-BB). Its horizontal profile, shown with a black curve in Fig.~\ref{fig:spectroscopy}b, matches very well with the envelope of the BQCL emission spectrum recorded with FTIR (red curve in Fig.~\ref{fig:spectroscopy}b).

The emission band of the chosen BQCL device ($1350-1600~\text{cm}^{-1}$) covers a spectral region extremely rich in signatures of vibrational modes of organic molecules, including vinyl C-H bends, aromatic ring stretch modes, carboxyl group modes, and more\cite{coates2000interpretation}. The transmission configuration described above can be used to detect the absorption spectra of thick layers of analytes. We showcase this for films of Polystyrene and PMMA spin-coated on bare CaF$_2$ chips (see Methods for details). Red curves in Fig.~\ref{fig:spectroscopy}d,e indicate the normalized intensity profiles of BQCL transmission images and are plotted together with the absorption spectra of the same films measured with FTIR (black curves). The profiles, normalized to the transmission of a bare CaF$_2$ chip, closely reproduce the reference data in both the amplitude and spectral position of the peaks. However, they manifest larger noise, which we attribute mainly to the room-temperature imaging camera utilized in our setup\cite{Cargioli2024Mar}. The increase of noise towards the edges of the spectral range is due to smaller laser intensity and thus larger contribution of the detector noise to the normalized signal. 

\begin{figure}[ht!]
    \centering
    \includegraphics[width=\textwidth]{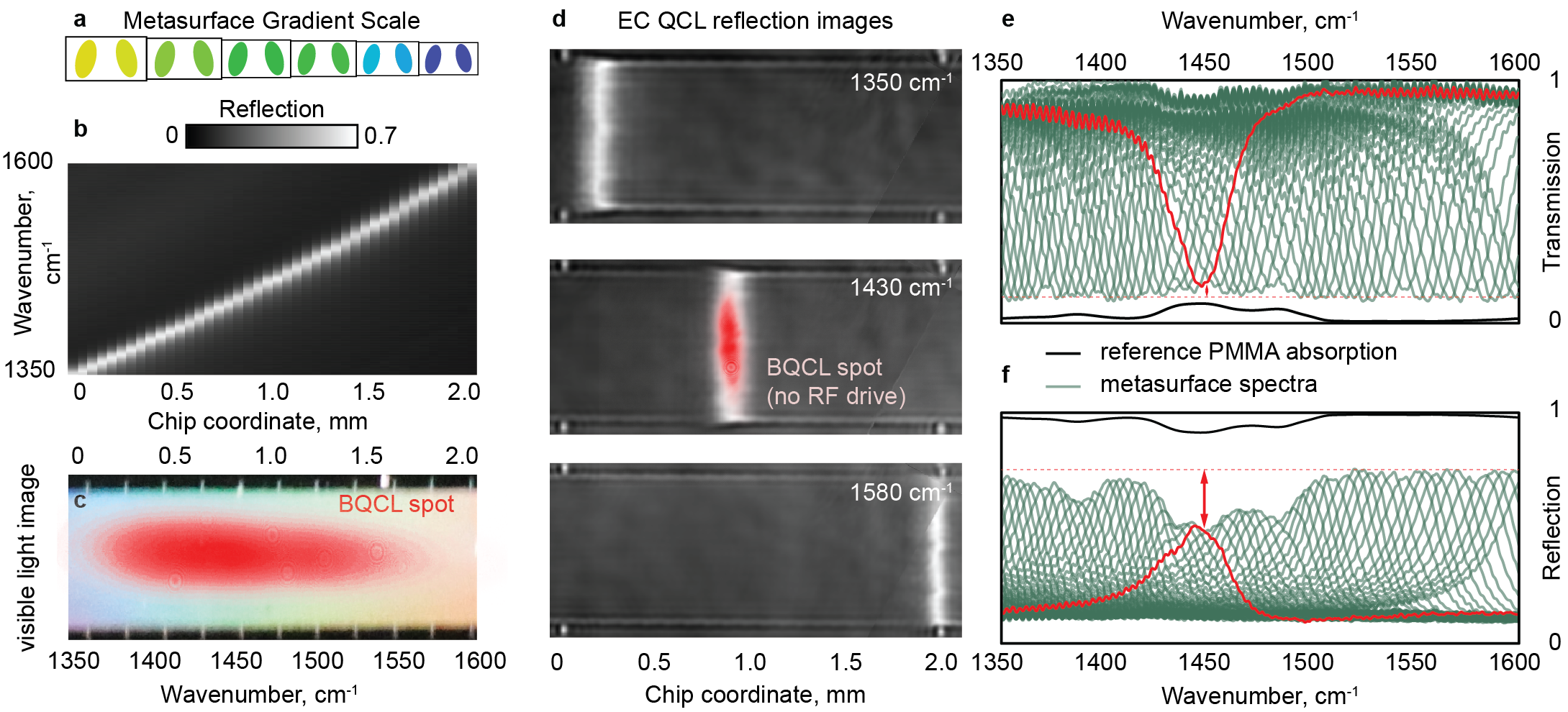}

    \caption{\textbf{Broadband gradient metasurface for imaging-based SEIRA spectroscopy} (a) Scheme of the unit cells of the resonance gradient metasurface consisting of two tilted Ge ellipses on CaF$_2$ substrate scaled along one of the chip coordinates. (b) Map of the reflection spectra of the gradient metasurface showing the evolution of the resonant peak with the chip coordinate. (c) Optical image of the gradient metasurface chip. Top axis shows the physical chip coordinate, bottom axis - the wavenumber of the resonant peak at the corresponding location on the chip. The red overlay shows the BQCL spot imaged with a mid-IR camera with the same scale (also shown in Fig.~\ref{fig:spectroscopy}c). (d) Reflection images of the gradient metasurface chip measured at different excitation frequencies with a mid-IR microscopy using external cavity tunable QCLs as its light source. The overlay shows the intensity profile of the BQCL spot focused on the sample surface in single-frequency emission regime (no RF drive). (e) Stack of FTIR transmission spectra of the broadband resonance gradient metasurface covered with PMMA measured at different chip coordinates. Polymer absorption has minor contribution to the spectra (indicated with small red arrow). (f) Stack of FTIR reflection spectra of the same chip. The amplitude of the peaks is strongly modulated by PMMA absorption (amplitude indicated with red arrow). Reference PMMA absorption profile is shown with black curves in both panels (inverted in panel f for convenience).}
    \label{fig:metasurface}
\end{figure}

To demonstrate imaging SEIRA spectroscopy with BQCL, we utilized resonant gradient metasurfaces, which allowed us to leverage the enhanced electromagnetic fields in the full spectral range of the laser emission. The metasurface design is based on elliptical Ge nanoresonators on CaF$_2$ substrate (Fig.~\ref{fig:Intro}
, and the scale of the unit cell of the metasurface is smoothly varied along one of the directions of the chip as illustrated in Fig.~\ref{fig:metasurface}a. The map of the reflectivity spectra of the fabricated chip measured with FTIR depending on the coordinate is shown in Fig.~\ref{fig:metasurface}b, with the resonance tuning range fully covering the BQCL emission band. We optimize the gradient to match the local resonance with the projected size of the BQCL spot on the sample surface as well as the dispersion of the grating (Fig.~\ref{fig:metasurface}c
). As compared to other mid-IR spectroscopy techniques, such as FTIR or tunable QCL spectroscopy, our approach provides more efficient utilization of light energy, as at each emission frequency all of the radiation is channeled towards the location on the sample that provides its resonant enhancement. We also ensure that the spatial broadening of the resonance, defined by the resonance quality factor as well as the speed of the gradient tuning, is matched with the size of the diffraction-limited laser spot size at a single frequency. This is illustrated in Fig.~\ref{fig:metasurface}d, which shows the reflection images of the gradient metasurface measured with a tunable QCL microscopy setup (Daylight Solutions Spero) at three different excitation frequencies. The semi-transparent overlay in the same panel shows the intensity profile of the laser spot on the chip surface in the single-frequency emission regime (no RF drive). The spatial scales of the images and the overlay are matched with the help of auxiliary labels fabricated on the chip. 

\begin{figure}[ht!]
    \centering
    \includegraphics[width=0.9\textwidth]{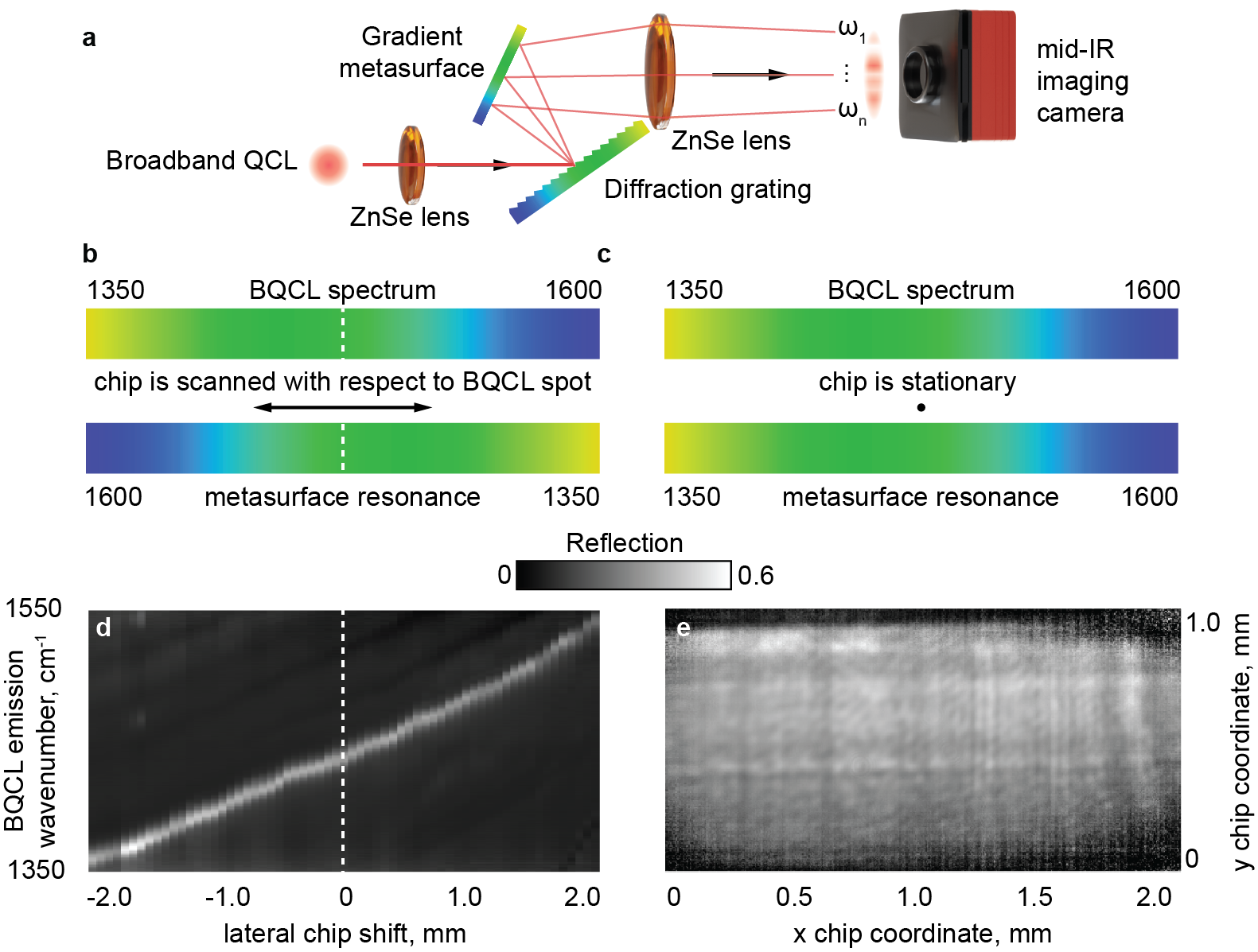}
    \caption{\textbf{Calibration of the alignment of gradient metasurfaces with broadband QCL emission for rapid SEIRA-based imaging spectroscopy.} (a) Scheme of BQCL imaging spectroscopy integrated with  a gradient metasurface. Spatial distribution of the graident metasurface resonance matches the dispersion of the BQCL beam wavelength after the diffraction grating.  (b,c) Schematics of calibration configurations for the imaging spectroscopy system. (b) Mismatched configuration: the dispersions of the BQCL spot and the metasurface resonance have the opposite sign. When the gradient metasurface is scanned along the BQCL spot, light is reflected only from the area where the dispersions intersect (marked with white dashed line). (c) Matched configuration: each frequency component of the BQCL addresses the location on the gradient with the corresponding resonance frequency. (d) Calibration data for mismatched configuration. Each vertical section of the map represents the profile of the BQCL spot for a certain shift of the chip from the spot center. (e) Calibration data for matched configuration. Image of the BQCL spot reflected from the gradient chip demonstrates uniformly high reflectance due to matching of the laser frequency components with local metasurface resonance.}
    \label{fig:calibration}
\end{figure}

Unlike the mid-IR absorption spectra of analytes on CaF$_2$ chips that were detected in transmission, the metasurface-enhanced absorption signatures manifest much stronger in the reflection spectra. Figure~\ref{fig:metasurface}e,f shows the comparison of transmission and reflection spectra measured from the gradient chip covered with a thin layer of PMMA. The amplitudes of the transmission dips (Fig.~\ref{fig:metasurface}e) and reflection peaks (Fig.~\ref{fig:metasurface}f) are modulated by the absorption of PMMA, its reference spectrum shown with black curves in both panels. The modulation magnitude (indicated with red arrows in Fig.~\ref{fig:metasurface}e,f) is almost six times stronger for reflection. Therefore, to fully benefit from the absorption signal enhancement, we changed the detection scheme to reflection as illustrated in Fig.~\ref{fig:calibration}a.

This new reflection configuration requires a thorough calibration to ensure that the resonance wavenumber is perfectly matched with the BQCL spectral component along the entire laser spot. For that, we performed two series of measurements illustrated in Fig.~\ref{fig:calibration}b,c. First, we positioned the chip so that dispersion of the resonance in the gradient is counteralinged with the dispersion of the frequency in the laser beam projected on the chip surface (Fig.~\ref{fig:calibration}b). In this configuration, the reflected image only manifests a narrow peak at the position on the chip where the dispersions cross each other (that is, the wavenumber of the local resonance is matched with the spectral component of the laser emission addressing this location). By scanning the chip along the laser spot, we observed the shift of the reflected peak in the recorded mid-IR image and correlated the wavenumber with the coordinates in the image using the known spatial dependence of the resonance in the gradient. Each vertical section of Fig.~\ref{fig:calibration}d represents an averaged intensity profile of the mid-IR reflection image for the corresponding shift of the gradient metasurface with respect to the projected laser spot. It demonstrates almost linear dependence of the wavenumber on the image coordinate 
. The measured shift distance allowed us to precisely calibrate the distance between the chip and the diffraction grating that ensures perfect matching of the metasurface size with the laser spot size. To verify the optimal match of both sizes and dispersions of the gradient and the laser spot, we then rotated the chip 180 degrees, switching to the aligned configuration (Fig.~\ref{fig:calibration}c). The reflection image recorded from the gradient normalized to the reflection image from the gold film (Fig.~\ref{fig:calibration}e) shows almost constant reflection within the laser spot, which confirms good matching of dispersions and, consequently, efficient local field enhancement for the entire available spectral range.

\begin{figure}[ht!]
    \centering
    \includegraphics[width=\textwidth]{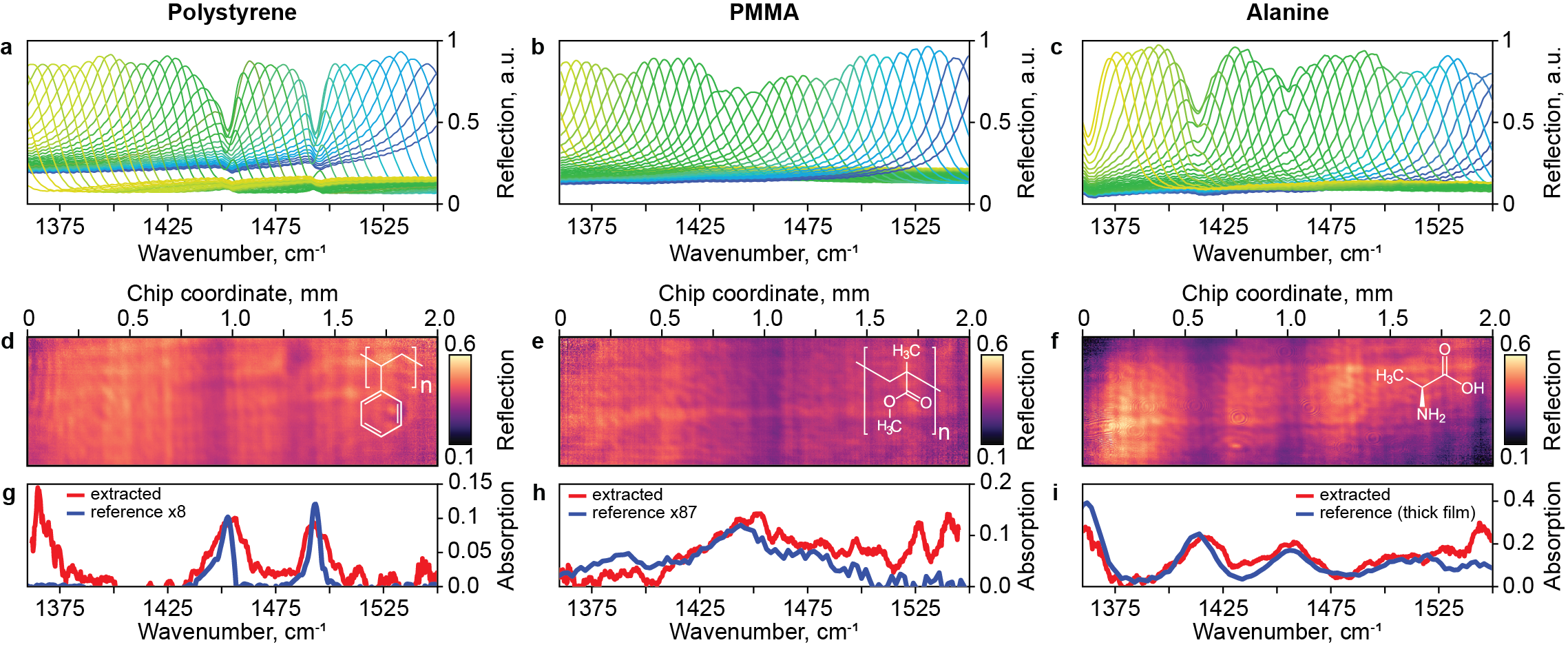}
    \caption{\textbf{Demonstration of the rapid single-frame imaging-based SEIRA of analytes with broadband QCL and gradient metasurfaces} (a-c) FTIR reflection spectra of gradient metasurface covered with layers of different analytes: (a) 300~nm polystyrene film, (b) 80~nm PMMA film, and (c) 45~nm alanine film. The envelope of the spectra stack measured at different locations along the gradient shows the absorption profiles of the analytes. (d-f) Images of the BQCL spots reflected from the gradient metasurface covered with the respective analytes. The images are normalized to reflection image from gold film measured in the same configuration. (g-i) Red curves: enhanced absorption spectra extracted from the intensity profiles of single-frame reflection images of the gradient metasurface. Blue curves: scaled reference absorption spectra extracted from reflection spectra of analytes on gold measured with FTIR.}
    \label{fig:sensing}
\end{figure}

Finally, we demonstrate the synergy of our BQCL imaging spectroscopy and gradient metasurfaces for enhanced detection of thin layers of organic molecules. We performed measurements for the gradient chip covered with analytes: polystyrene, PMMA, and L-alanine (see Methods for details on the coating routines). The absorption spectra of these three analytes reveal distinctively different signatures attributed to their respective vibrational bands \cite{Socrates2004Jun}. The absorption peaks can be seen in the envelope of the reflection spectra measured with FTIR along the length of the gradient metasurface covered with the respective analyte (Fig.~\ref{fig:sensing}a-c). The common feature of all three compounds is the peak at $\approx$1450~$\text{cm}^{-1}$, which corresponds to $\text{CH}_{2}$ scissor vibrations. The unique feature of polystyrene (Fig.~\ref{fig:sensing}a) is the contribution from aromatic ring stretch modes that give rise to two pronounced peaks at 1450 and 1490~cm$^{-1}$. PMMA spectra (Fig.~\ref{fig:sensing}b) are mostly defined by C-H and CH$_2$ vibrations. Finally, the distinctive features of alanine spectra (Fig.~\ref{fig:sensing}c) are the carboxylate COO- symmetric stretch mode at 1410~$\text{cm}^{-1}$ and O-H deformation vibrations below 1380~$\text{cm}^{-1}$, as well as the onset of amide-II vibration band towards 1550~$\text{cm}^{-1}$.  

As revealed by the FTIR spectra, the presence of analytes modifies the reflectance of the gradient metasurface at the locations where the metasurface resonance matches the absorption peaks. Illumination of the gradient chip with the diffraction spot of our BQCL converts this spatially encoded absorption into a ``barcode'' image defined by the absorption signature of the corresponding molecule. This molecular absorption barcode can be captured in a single frame of a mid-IR imaging camera, providing rapid identification of the analyte. The images captured for three analytes and normalized to reflection from gold film are shown in Fig.~\ref{fig:sensing}d-f. One can immediately see pronounced reflection minima corresponding to the absorption resonances of the respective molecules that manifest as dark vertical stripes in the images. The line defects visible in the images come from the stitching of the write fields of the e-beam tool. However, these artifacts do not interfere with the strongly pronounced absorption features. 

The interaction of molecular vibrations with the resonant modes of the gradient metasurface, which is the essence of SEIRA, is crucial for resolving the weak absorption signatures of thin molecular layers. However, because of the spatial localization of the enhanced optical field, the actual enhancement of the extracted absorption signal decreases for thicker layers of analyte that extend beyond the mode volume. Accordingly, we used different thicknesses of molecular layers for extraction of metasurface-enhanced absorption barcodes and compared their profile with reference absorption data extracted from  FTIR reflectance spectra of films deposited on gold (see Methods for details on the measurement configuration). For polystyrene, which has the weakest absorption signature in the available wavelength range, we used the thickest film, measured at 300~nm according to AFM data. The absorption spectrum of the analyte extracted from the molecular barcode image and the FTIR reference measured from a film spin-coated on gold at the same conditions are shown in Fig.~\ref{fig:sensing}g. Due to the large thickness of the film, the enhancement is only 8. A thinner ($\approx$ 80~nm) PMMA film demonstrates almost the same amplitude of modulation of QCL reflectance profile for considerably weaker FTIR absorption readings, which converts to an estimated enhancement factor of 87 (Fig.~\ref{fig:sensing}b). Interestingly, the thinnest (45~nm) alanine film demonstrates the strongest modulation amplitude, while the absorption is not resolved at all in the FTIR reflection spectra (in Fig.~\ref{fig:sensing}i, we used reflection data from a grazing angle objective as a reference instead). This can be explained by the different deposition methods (spin-coating for polymer films vs. thermal evaporation for alanine) which lead to different overlaps of the analytes with the optical mode volume. 

In addition to the strong absorption signal enhancement, our platform provides unparalleled acquisition speed. All presented imaging SEIRA data are captured at 7~fps readout (14~ms  per frame). This represents a 3~orders of magnitude improvement over the FTIR measurements of thin films without SEIRA enhancement (blue lines in Fig.~\ref{fig:sensing}g-i, 120s per measurement) and up to 4 orders improvement as compared to FTIR SEIRA measurements that also involve mechanical scanning of the collection area along the gradient chip (Fig.~\ref{fig:sensing}a-c, 20~min per measurement). This represents a breakthrough advancement for high-throughput, highly sensitive molecular diagnostics in mid-IR.

\section*{Discussion}

To conclude, in this paper we demonstrated rapid imaging-based mid-IR spectroscopy enabled by the synergy of broadband resonance gradient metasurfaces and a quantum cascade laser generating remarkably broad instantaneous emission spectrum (250~cm$^{-1}$).  As we match the dispersion of the BQCL spot projected on the metasurface with the spatial gradient of the resonance, each frequency component of the emission exclusively addresses the location on the chip that provides its resonant enhancement. This maximizes the fraction of BQCL power that experiences the enhanced interaction with IR vibrational modes via the metasurface resonances and enables high-throughput acquisition of the metasurface-enhanced molecular absorption signatures in a single frame of a room-temperature mid-IR camera in the form of a barcode-like image. We showcased the concept by detecting unique absorption fingerprints of thin layers of three different molecules deposited on the gradient metasurface with a footprint of less than 1~mm$^2$ with sub-second acquisition time that is up to 3 orders of magnitude shorter than for FTIR and external cavity QCL based spectroscopy techniques. We envision that our approach will facilitate rapid and highly specific molecular diagnostics across a wide range of chemical and biological applications and inspire the development of highly compact and miniaturized mid-IR bioanalytical devices.

\section*{Methods}
\subsection*{Quantum cascade laser fabrication and operation}
The QCL, which embeds an heterogeneous active region in order to maximize the emission spectral region, has been processed through a buried heterostructure technique and mounted eptiaxial side up. The laser is wire-bonded to an SMA connector which is used to drive both the DC current and the modulation signal at the same time via the use of a high power bias-tee. The RF signal has a power of 29~dBm and a frequency of 500~MHz. The DC operation point is typically chosen to be close to the threshold value to profit from the most stable spectra. 

\subsection*{Metasurface fabrication}
Gradient metasurfaces based on germanium (\text{Ge}) resonators were fabricated on calcium difluoride (\(\text{CaF}_2\)) substrates. First, the material stack was prepared by subsequent sputtering of a 5~nm silicon oxide (\(\text{SiO}_2\)) adhesion layer and a 700~nm Ge layer on top of a 1~mm thick CaF$_2$ chip. To exclude the absorption from the alumina hard mask that is usually used for patterning of Ge films, we used inverted metasurface pattern, written using electron beam lithography (Raith EPBG500+) on a spin-coated single-layer PMMA (PMMA 495k A8, 4000 rpm) positive tone resist. After the development, the pattern was directly transferred to the Ge film using a fluorine-based dry plasma etching process (Alcatel AMS 200 SE). To suppress the defects arising from stitching errors of the neighboring write fields of the e-beam tool, we additionally customized the e-beam exposure sequence to follow a meander path in  column-by-column fashion.

\subsection*{Deposition of molecular films}

Layers of PMMA and polystyrene were deposited on clean CaF$_2$ chips and gradient metasurface using spin-coating. We used different solvents for the two polymers: anisole for PMMA and o-xylene for polystyrene to ensure good solubility even for high concentrations. For transmission tests, 8wt\% PMMA and 6wt\% polystyrene solutions were spin-coated at 2000~rpm on CaF$_2$ chips, which yielded thick films with easily resolved absorption. The thickness was measured with an optical profilometer, which showed 700~nm and 550~nm for PMMA and polystyrene, respectively. For metasurface-enhanced absorption measurements, we used thinner films, with 2wt\% PMMA spin-coated at 2000~rpm and 6wt\% PS spin-coated at 3000 rpm, which resulted in a film thickness of 80~nm and 300~nm, respectively. It should be noted, however, that the profilometry data from films spin-coated on flat surfaces give only a rough estimate of the amount of material deposited on the metasurface, as its topography may modify the local film thickness substantially. From that perspective, data obtained from evaporation of alanine gives a more precise estimation of the amount of detected analyte.

Deposition of alanine (Thermo Scientific) was performed using thermal evaporation in a VacoTec vacuum chamber. The distance between the tungsten boat for Joule-effect evaporation and the target chip was $\approx$ 40~cm. The evaporator chamber was evacuated to a pressure of $10^{-5}$~mbar. Electrical current through the boat filled with $\approx$ 100~mg of alanine was used to maintain steady evaporation rate (1-1.2~\AA/s) monitored using a quartz crystal microbalance. The readings of the thickness monitor were additionally calibrated using atomic force microscopy. 

\subsection*{Imaging spectroscopy measurements}
The custom QCL chip sustained at -20$^{\circ}$C with a Peltier element was driven by a DC current source at 220~mA (PpqSense QubeCL) with an additional RF signal superimposed using a bias tee. 500~MHz RF signal was provided by Rohde\&Schwarz SMB100B signal generator at a base amplitude of -26~dBm and amplified with a broadband high power amplifier module (RFlambda RFLUPA5M25MK) up to 29~dBm. Laser light was focussed on the chip surface with a 4~cm ZnSe lens (Thorlabs) through a 75~g/mm diffraction grating (Thorlabs GR2550-07106). The images were captured with a room-temperature microbolometer array camera (DataRay WinCamD-IR-BB) with 640$\times$480 pixels.



\section*{Acknowledgements}
We acknowledge the European Union's Horizon Europe Research and Innovation Programme under agreements 101046424 (TwistedNano) and 101070700 (MIRAQLS). This work was supported by the Swiss State Secretariat for Education, Research and Innovation (SERI) under contract numbers 22.00018 and 22.00081. The authors thank Alpes Lasers for lending the broadband QCL chip. The authors acknowledge the use of nanofabrication facilities at the Center of MicroNano Technology of \'Ecole Polytechnique F\'ed\'erale de Lausanne.

\section*{Author Contributions Statement}
I.S. fabricated the metasurface chips and performed the measurements with help of F.R. A.C. and D.P. fabricated and tested the quantum cascade laser chip and performed its integration in the measurement setup. I.S and F.R. performed the numerical simulations. J.F. and H.A. conceived the idea and guided the research.

\section*{Competing Interests Statement}
The authors declare no competing interests.

\end{document}